\documentclass[cpp,fleqn]{w-art}
\usepackage{times}
\usepackage{w-thm}

\usepackage[]{graphicx}
\usepackage[breaklinks=true,colorlinks=true,linkcolor=blue,urlcolor=blue,citecolor=blue]{hyperref}

\usepackage{xcolor}    
\begin{document}
\Volume{42}
\Issue{1}
\Month{01}
\Year{2003}



\title[Free Energy of the Uniform Electron Gas]{Free Energy of the Uniform Electron Gas: Testing Analytical Models against First Principle Results\footnote{Dedicated to Werner Ebeling on the occasion of his 80th birthday.}}


\author{Simon Groth\footnote{Corresponding
     author: e-mail: {\sf groth@theo-physik.uni-kiel.de}}} \address[]{Institut f\"ur Theoretische Physik und Astrophysik, Christian-Albrechts-Universit\"{a}t zu Kiel, D-24098 Kiel, Germany}
\author{Tobias Dornheim}
\author{Michael Bonitz}
\begin{abstract}
The uniform electron gas is a key model system in the description of matter, including dense plasmas and solid state systems. However, the simultaneous occurence of quantum, correlation, and thermal effects makes the theoretical description challenging. For these reasons, over the last half century many analytical approaches have been developed the accuracy of which has remained unclear. We have recently obtained the first \textit{ab initio} data for the exchange correlation free  energy of the uniform electron gas [T. Dornheim \textit{et al.}, Phys.~Rev.~Lett.~\textbf{117}, 156403 (2016)] which now provides the opportunity to assess the quality of the mentioned approaches and  parametrizations. Particular emphasis is put on the warm dense matter regime, where we find significant discrepancies between the different approaches.
\end{abstract}
\maketitle                   





\section{Introduction}

Over the last decade there has emerged growing interest in the  so-called warm dense matter (WDM), which is of key importance for the description of, e.g., astrophysical systems \cite{knudson,militzer}, laser-excited solids \cite{ernst}, and inertial confinement fusion targets \cite{nora,schmit,hurricane3}.
The WDM regime is characterized by the simultaneous occurence of strong (moderate) correlations of ions (electrons), thermal effects as well as quantum effects of the electrons. In dimensionless units, typical parameters are the Brueckner parameter $r_s=\overline{r}/a_\textnormal{B}$ and the reduced temperature $\theta=k_\textnormal{B}T/E_\textnormal{F}$, both being of the order of unity (more generally in the range $0.1 \dots 10$). Here $\overline{r}$ and $a_\textnormal{B}$ denote the mean interparticle distance and the Bohr radius, respectively. A third relevant parameter is the classical coupling parameter of the ionic component, $\Gamma_i = Z_i^2e^2/\overline{r}k_BT$ which is often larger than unity indicating that the ionic component is far from an ideal gas. 
This makes the theoretical description of this peculiar state of matter particularly challenging, as there is no small parameter to perform an expansion around. 

In the ground state, there exists a large toolkit of approaches that allow for the accurate description of manifold physical systems, the most successful of which arguably being Kohn-Sham density functional theory (DFT), e.g. \cite{ks,dft_review}.
The basic idea of DFT is to map the complicated and computationally demanding quantum many-body problem onto an effective single-particle problem. This would be exact if the correct exchange-correlation functional of the system of interest was available which is, of course, not the case. In practice, therefore, one has to use an approximation. The foundation of the great success of DFT has been the local density approximation (LDA), i.e., the usage of the exchange-correlation energy $E_{xc}$ of the uniform electron gas (UEG) with the same density as the more complicated system of interest. Accurate data for $E_{xc}$ of the UEG was obtained by Ceperley and Alder \cite{alder} using a Quantum Monte Carlo (QMC) method, from which Perdew and Zunger \cite{perdew} constructed a simple parametrization with respect to density, $E_{xc}(r_s)$, that is still used to this day.

However, the accurate description of warm dense matter requires to extend DFT to finite temperature. This has been realized long ago  by Mermin \cite{mermin} who used a superposition of excited states weighted with their thermal occupation probability. A strict approach to the thermodynamic properties of this system also requires an appropriate finite-temperature extension of the LDA, in particular replacement of the ground state energy functionals 
by free energies, i.e. $E \to f = E - TS$. This means, a finite-temperature version of the LDA requires accurate parametrizations of the  {\em exchange correlation free energy} with respect to temperature and density \cite{karasiev2,dharma, dharma_cpp15, gga,burke,burke2}, i.e., $f_{xc}(r_s,\theta)$, even though in some cases the entropic correction maybe small. This seemingly benign task, however, turns out to be far from trivial because accurate date for the free energy are much more involved than the ground state results. While for the ground state reliable QMC data are known for a long time, until recently \cite{tim2,tim_prl,dornheim,dornheim2,groth,dornheim3,dornheim_cpp,malone,malone2,dornheim_prl,dornheim_pop_17}, the notorious fermion sign problem \cite{loh,troyer} has prevented reliable QMC simulations in the warm dense regime. 
Therefore, during the recent four decades many theoretical approaches to $f_{xc}(r_s,\theta)$ have been developed that lead to a variety of parametrizations, for an overview on early works, see e.g.~\cite{kraeft-book, dewitt-tribute} . Some of them have gained high popularity and they were successfully applied in many fields, even though there accuracy has remained poorly tested. It is the purpose of the present article to present such a quantitative comparison of earlier models to new simulation results.

In Sec.~\ref{sec:free}, we introduce a selection of such functions. First, we analyze the purely analytical expression presented  by Ebeling and co-workers, e.g. \cite{ebeling1}. Next, we study  functional fits to linear response data based on static local field correction schemes that were suggested by Singwi, Tosi, Land, and Sj\"olander (STLS) \cite{stls_original} (Sec.~\ref{sub:ichimaru}) and Vashishta and Singwi (VS) \cite{vs_original} (Sec.~\ref{sub:vs}). As a fourth example we  consider the quantum-classical mapping developed by Dharma-wardana and Perrot (PDW) \cite{pdw_prl,pdw}  (Sec.~\ref{sub:pdw}).
Finally, we consider the recent parametrization by Karasiev \textit{et al.}~(KSDT) \cite{karasiev} (Sec.~\ref{sub:ksdt}) that is based on the QMC data by Brown \textit{et al.}~that became available recently~\cite{brown}. However, those data have a limited accuracy due to (i) the usage of the fixed node approximation \cite{node} and (ii) an inappropriate finite-size correction (see \cite{dornheim_prl}) giving rise to systematic errors in the free energy results as we will show below.
In Sec.~\ref{sec:results}, we compare all aforementioned parametrizations of $f_{xc}$ to the new, accurate QMC data by Dornheim \textit{et al.}~\cite{dornheim_prl} that are free from any systematic bias, and, hence, allow us to gauge the accuracy of models. A particular emphasis is laid on the warm dense matter regime.

\section{\label{sec:free}Free Energy Parametrizations}

\subsection{\label{sub:ebeling}Ebeling's Pad\'e formulae}
The idea to produce an analytical formula for the thermodynamic quantities that connect known analyitical limits via a smooth Pad\'e approximant is due to Ebeling, Kraeft, Richert and co-workers \cite{ebeling_richert81, ebeling_ebeling_richert82, ebeling_richert85_1, ebeling_richert85_2}. These approximations have been quite influential in the description of nonideal plasmas and electron-hole plasmas in the 1980s and 1990s receiving, in part, a substantial amount of citations.
These approximations have been improved continuously in the following years, and we, therefore, only discuss the more recent versions, cf.~\cite{ebeling1,ebeling2} and references therein. 

Ebeling {\em et al.} used Rydberg atomic units and introduced a reduced thermal density
\begin{equation}
\overline{n} = n\Lambda^3 = 6\sqrt{\pi}r_s^{-3}\tau^{-3/2} \quad ,
\end{equation}
with the usual thermal wavelength $\Lambda$ and $\tau=k_\textnormal{B}T/\textnormal{Ry}$ being the temperature in energy units.
The Pad\'e approximation for $f_{xc}$ then reads \cite{ebeling1}
\begin{eqnarray}
f_{xc}^\textnormal{Ebeling,Ry}(r_s,\tau) = - \frac{ f_0(\tau) \overline{n}^{1/2} + f_3(\tau) \overline{n} + f_2 \overline{n}^2\epsilon^\textnormal{Ry}(r_s) }{ 1 + f_1(\tau) \overline{n}^{1/2} + f_2 \overline{n}^2 } \quad , \label{eq:ebeling_Ry}
\end{eqnarray}
with the coefficients
\begin{eqnarray}
f_0(\tau) = \frac{2}{3}\left(\frac{\tau}{\pi}\right)^{1/4}\ ,\quad f_1(\tau) = \frac{1}{8f_0(\tau)}\sqrt{2}(1+\textnormal{log}(2))\ ,\quad f_2 = 3 \ ,\quad f_3(\tau) = \frac{1}{4}\left( \frac{\tau}{\pi}\right)^{1/2}\ ,
\end{eqnarray}
and the ground state parametrization for the exchange correlation energy
\begin{equation}
\epsilon^\textnormal{Ry}(r_s) = \frac{0.9163}{r_s} + 0.1244\ \textnormal{log}\left( 1 + \frac{ 2.117 r_s^{-1/2}}{1+0.3008\sqrt{r_s}}\right) \quad .
\end{equation}
To achieve a better comparability with the other formulas discussed below, we re-express Eq.~(\ref{eq:ebeling_Ry}) in Hartree atomic units as a function of $r_s$ and the reduced temperature $\theta=k_\textnormal{B}T/E_\textnormal{F}$:
\begin{eqnarray}
\label{eq:ebeling}f_{xc}^\textnormal{Ebeling,Ha}(r_s,\theta) &=& -\frac{1}{2} \frac{ Ar_s^{-1/2}\theta^{-1/2} + B r_s^{-1}\theta^{-1} + C\theta^{-3}\epsilon^\textnormal{Ry}(r_s) }{ 1 + D\theta^{-1}r_s^{1/2}+ C \theta^{-3}} \quad , \quad \textnormal{with} \\
A &=& \frac{2}{3\sqrt{\pi}}\left(\frac{8}{3}\right)^{1/2}\left(\frac{4}{9\pi}\right)^{-1/6}\ , \quad B = \frac{2}{3\pi}\left( \frac{4}{9\pi} \right)^{-1/3}\ , \quad C=\frac{64}{3\pi}\ , \quad \\ \nonumber D &=& \frac{ (1+\textnormal{log}(2))\sqrt{3} }{4}\left( \frac{4}{9\pi}\right)^{1/6}\ .
\end{eqnarray}
Evidently, Eq.~(\ref{eq:ebeling}) incorporates the correct ground state limit
\begin{equation}
\lim_{\theta \to 0} f_{xc}^\textnormal{Ebeling,Ha}(r_s,\theta) = -\frac{1}{2}\epsilon^\textnormal{Ry}(r_s) \quad ,
\end{equation}
where the pre-factor $1/2$ is due to the conversion between Rydberg and Hartree units. Similarly, in the high-temperature limit the well-known Debye-H\"uckel result is recovered, e.g.~\cite{dewitt}:
\begin{eqnarray}\label{eq:DBH}
\lim_{\theta \to \infty} f_{xc}^\textnormal{Ebeling,Ha}(r_s,\theta) = -\frac{1}{2} A\ r_s^{-1/2} \theta^{-1/2} 
= - \frac{1}{\sqrt{3}}r_s^{-3/2}T^{-1/2} \quad .
\end{eqnarray}
Results for the warm dense UEG computed from these formulas are included in the figures below. For Pad\'e approximations to the UEG at strong coupling in the quasi-classical regime, see, e.g., Ref.~\cite{stolz_ebeling}.

\subsection{\label{sub:ichimaru}Parametrization by Ichimaru and co-workers}
In the mid-eighties, Tanaka, Ichimaru and co-workers \cite{stls85,stls} extended the original STLS scheme \cite{stls_original} for the static local field corrections to finite temperature and numerically obtained the interaction energy $V$ (per particle) of the UEG via integration of the static structure factor $S(k)$,
\begin{equation}\label{eq:V}
V = \frac{1}{2} \int_{k<\infty} \frac{ \textnormal{d}\mathbf{k} }{ (2\pi)^3 } [S(\mathbf{k}) - 1]\frac{4\pi}{\mathbf{k}^2} \quad ,
\end{equation}
for $70$ parameter combinations with $\theta=0.1,1,5$ and $r_s\sim10^{-3},\dots,74$.
Subsequently, there has been introduced a parametrization for $V$ as a function of $r_s$ and $\theta$ \cite{ichimaru_rev_2,ichimaru_rev}
\begin{equation}\label{eq:ichi_V}
V(r_s,\theta) = - \frac{1}{r_s} \frac{ a_\textnormal{HF}(\theta) + \sqrt{2}\lambda r_s^{1/2} \textnormal{tanh}(\theta^{-1/2}) B(\theta) + 2\lambda^2r_sC(\theta)E(\theta)\textnormal{tanh}(\theta^{-1}) }{1+ \sqrt{2}\lambda r_s^{1/2}D(\theta)\textnormal{tanh}(\theta^{-1/2}) + 2\lambda^2 r_s E(\theta)}\ ,
\end{equation}
with the definitions
\begin{eqnarray}\label{eq:ichi_definitions}
a_\textnormal{HF}(\theta) &=& 0.610887\ \textnormal{tanh}\left(\theta^{-1}\right) \frac{ 0.75 + 3.4363\theta^2 -0.09227\theta^3 +  1.7035\theta^4 }{ 1+ 8.31051\theta^2 + 5.1105\theta^4 }\\
B(\theta) &=& \frac{  x_1 + x_2\theta^2 + x_3 \theta^4 }{ 1 + x_4\theta^2 + x_5\theta^4 }\ , \quad 
C(\theta) = x_6 + x_7 \textnormal{exp}\left( -\theta^{-1} \right) \ , \\
D(\theta) &=& \frac{ x_8 + x_9\theta^2 + x_{10}\theta^4 }{ 1 + x_{11}\theta^2 + x_{12}\theta^4} \ , \quad 
E(\theta) = \frac{x_{13}+x_{14}\theta^2+x_{15}\theta^4}{1+x_{16}\theta^2 + x_{17}\theta^4} \quad .
\end{eqnarray}
In addition to the exact limits for $\theta\to 0$ and $\theta\to\infty$, the parametrization from Eq.~(\ref{eq:ichi_V}) also approaches the well-known Hartree-Fock limit for high density,
\begin{equation}
\lim_{r_s\to0}V(r_s,\theta) = - \frac{ a_\textnormal{HF}(\theta)}{r_s} \quad ,
\end{equation}
which has been parameterized  by Perrot and Dharma-wardana \cite{pdw84}, see Eq.~(\ref{eq:ichi_definitions}).
Naturally, the free parameters $x_i$, $i=1,\dots,17$ have been determined by a fit of Eq.~(\ref{eq:ichi_V}) to the STLS data for $V$ and the resulting values are listed in Tab.~\ref{tab:ichi}.
From the interaction energy $V(r_s,\theta)$, the free exchange-correlation energy is obtained by integration
\begin{equation}\label{eq:fxc}
f_{xc}(r_s,\theta) = \frac{1}{r_s^2}\int_0^{r_s} \textnormal{d}\overline{r}_s\ \overline{r}_s  V(\overline{r}_s,\theta) \quad .
\end{equation}
Plugging in the expression for $V(r_s,\theta)$ from Eq.~(\ref{eq:ichi_V}) into (\ref{eq:fxc}) gives the final parametrization for $f_{xc}(r_s,\theta)$
\begin{eqnarray}\label{eq:ichi_fxc}
f_{xc}(r_s,\theta) = &-& \frac{1}{r_s}\frac{c(\theta)}{e(\theta)} \\ \nonumber
&-& \frac{ \theta}{2 e(\theta) r_s^2\lambda^2 } \left[ \left( a_\textnormal{HF}(\theta)-\frac{c(\theta)}{e(\theta)}\right)
-\frac{d(\theta)}{e(\theta)}\left( b(\theta) - \frac{ c(\theta)d(\theta)}{e(\theta)}\right)\right] \\ \nonumber
&\times&\textnormal{log}\left|  \frac{ 2 e(\theta) \lambda^2 r_s }{ \theta} + \sqrt{2}d(\theta)\lambda r_s^{1/2} \theta^{-1/2} +1 \right| \\ \nonumber
&-& \frac{\sqrt{2}}{e(\theta)}\left( b(\theta) - \frac{ c(\theta)d(\theta) }{e(\theta)}\right) \frac{ \theta^{1/2} }{r_s^{1/2}\lambda}\\ \nonumber
&+& \frac{ \theta }{ r_s^2\lambda^2 e(\theta) \sqrt{4e(\theta)-d^2(\theta)}}\left[ d(\theta)\left(a_\textnormal{HF}(\theta)-\frac{c(\theta)}{e(\theta)}\right)\right. \\ \nonumber &+& \left. \left(2-\frac{d^2(\theta)}{e(\theta)}\right)\left(b(\theta)-\frac{c(\theta)d(\theta)}{e(\theta)}\right)\right] \\ \nonumber
&\times&\left[ \textnormal{atan}\left( \frac{ 2^{3/2} e(\theta) \lambda r_s^{1/2} \theta^{-1/2} + d(\theta) }{ \sqrt{4e(\theta)-d^2(\theta)} } \right) - \textnormal{atan}\left( \frac{ d(\theta) }{ \sqrt{4e(\theta)-d^2(\theta)}}\right)\right] \ ,
\end{eqnarray}
with the abbreviations
\begin{eqnarray}
b(\theta) &=& \theta^{1/2}\ \textnormal{tanh}\left( \theta^{-1/2} \right) B(\theta) \ , \quad
c(\theta) = C(\theta)e(\theta), \\ \nonumber 
d(\theta) &=& \theta^{1/2}\ \textnormal{tanh}\left( \theta^{-1/2} \right) D(\theta) \ , \quad 
e(\theta) = \theta\ \textnormal{tanh}\left( \theta^{-1} \right) E(\theta) \quad .
\end{eqnarray}

\begin{table}
\centering
\begin{tabular}{ c c c c c  }
$x_1$ & $x_2$ & $x_3$ &
$x_4$ &
$x_5$  \\ \hline
$3.4130800\times10^{-1}$ &
$1.2070873\times10$ &
$1.148889\times10^{0}$ &
$1.0495346\times10$ &
$1.326623\times10^0$\vspace{0.2cm} \\
$x_6$ & $x_7$ & $x_8$ &
$x_9$ &
$x_{10}$  \\ \hline
$8.72496\times10^{-1}$ &
$2.5248\times10^{-2}$ &
$6.14925\times10^{-1}$ &
$1.6996055\times10$ &
$1.489056\times10^0$ \vspace{0.2cm}\\
$x_{11}$ &
$x_{12}$ &
$x_{13}$ &
    $x_{14}$ &
    $x_{15}$ \\ \hline
    $1.010935\times10$ &
$1.22184\times10^0$ &
$5.39409\times10^{-1}$ &
$2.522206\times10^{0}$ &
$1.78484\times10^{-1}$ \vspace{0.2cm} \\
$x_{16}$ &
$x_{17}$ \\ \hline
$2.555501\times10^{0}$ &
$1.46319\times10^{-1}$

  \end{tabular}
\caption{Fit parameters by Ichimaru \cite{ichimaru_rev} for the $f_{xc}(r_s,\theta)$ parametrization from Eq.~(\ref{eq:ichi_fxc}), fitted to STLS data \cite{stls}.}
\label{tab:ichi}
\end{table}

\subsection{\label{sub:vs}Vashishta-Singwi parametrization}
Despite the overall good performance of STLS in the ground state \cite{bohm}, it has long been known that this scheme does not fulfill the compressibility sum-rule (CSR, see, e.g., Ref.~\cite{stls2} for a detailed discussion). To overcome this obstacle, Vashishta and Singwi \cite{vs_original} have introduced modified local field corrections (VS), where the CSR it automatically fulfilled. This idea had been extended in an approximate way to finite temperature by Stolzmann and R\"osler \cite{stolzmann}, until more recently Sjostrom and Dufty \cite{stls2} obtained an exhaustive data set of results that are exact within the VS framework.

As already explained in the previous section for the STLS data, they have first calculated the static structure factor $S(k)$, computed the interaction energy $V$ by integration (Eq.~(\ref{eq:V}), fitted the parametrization from Eq.~(\ref{eq:ichi_V}) to this data, and thereby obtained the desired parametrization of $f_{xc}(r_s,\theta)$ as given in Eq.~(\ref{eq:ichi_fxc}) (albeit with the new fit parameters listed in Tab.~\ref{tab:vs}).

\begin{table}
\centering
\begin{tabular}{ c c c c c  }
$x_1$ & $x_2$ & $x_3$ &
$x_4$ &
$x_5$  \\ \hline
$1.8871493\times10^{-1}$ &
$1.0684788\times10$ &
$1.1088191\times10^{2}$ &
$1.8015380\times10$ &
$1.2803540\times10^2$\vspace{0.2cm} \\
$x_6$ & $x_7$ & $x_8$ &
$x_9$ &
$x_{10}$  \\ \hline
$8.3331352\times10^{-1}$ &
$-1.1179213\times10^{-1}$ &
$6.1492503\times10^{-1}$ &
$1.6428929\times10$ &
$2.5963096\times10$ \vspace{0.2cm}\\
$x_{11}$ &
$x_{12}$ &
$x_{13}$ &
    $x_{14}$ &
    $x_{15}$ \\ \hline
    $1.0905162\times10$ &
$2.9942171\times10$ &
$5.3940898\times10^{-1}$ &
$5.8869626\times10^{4}$ &
$3.1165052\times10^{3}$ \vspace{0.2cm} \\
$x_{16}$ &
$x_{17}$ \\ \hline
$3.8887108\times10^{4}$ &
$2.1774472\times10^{3}$

  \end{tabular}
\caption{Fit parameters by Sjostrom and Dufty \cite{stls2} for the $f_{xc}(r_s,\theta)$ parametrization from Eq.~(\ref{eq:ichi_fxc}), fitted to VS data.}
\label{tab:vs}
\end{table}

\subsection{\label{sub:pdw}Perrot-Dharma-wardana parametrization}
Dharma-wardana and Perrot \cite{pdw_prl,pdw} have introduced an independent, completely different idea. In particular, they employ a \textit{classical mapping} such that the correlation energy of the electron gas at $T=0$ (that has long been known from QMC calculations~\cite{alder,perdew}) is exactly recovered by the simulation of a classical system at an effective ``quantum temperature'' $T_q$. However, due to the lack of accurate data at finite $T$, an exact mapping had not been possible, and the authors introduced a modified temperature $T_c$, where they assumed an interpolation between the exactly known ground state and classical (high $T$) regimes, $T_{c}=\sqrt{T^2+T^2_q}$. Naturally, at warm dense matter conditions this constitutes a largely uncontrolled approximation.

To obtain the desired parametrization for $f_{xc}$, extensive simulations of the UEG in the range of $r_s=1,\dots,10$ and $\theta=0,\dots,10$ were performed. These have been used as input for a fit with the functional form
\begin{eqnarray}\label{eq:pdw}
f_{xc}(r_s,\theta) &=& \frac{ \epsilon(r_s) - P_1(r_s,\theta) }{ P_2(r_s,\theta) }, \\ \nonumber
P_1(r_s,\theta) &=& \left(A_2(r_s)u_1(r_s) + A_3(r_s)u_2(r_s)\right) \theta^2 Q^2(r_s) + A_2(r_s)u_2(r_s)\theta^{5/2}Q^{5/2}(r_s), \\ \nonumber 
P_2(r_s,\theta) &=& 1 + A_1(r_s)\theta^2 Q^2(r_s) + A_3(r_s)\theta^{5/2}Q^{5/2}(r_s) + A_2(r_s)\theta^3Q^3(r_s), \\ \nonumber
Q(r_s) &=& \left( 2 r_s^2 \lambda^2 \right)^{-1}\ , \quad n(r_s) = \frac{3}{4\pi r_s^3}\ , \quad u_1(r_s) = \frac{\pi n(r_s)}{2}\ , \quad u_2(r_s) = \frac{2\sqrt{\pi n(r_s)}}{3}, \\ \nonumber
A_k(r_s) &=& \textnormal{exp}\left( \frac{ y_k(r_s) + \beta_k(r_s)z_k(r_s) }{ 1 + \beta_k(r_s) }
\right)\ , \quad \beta_k(r_s) = \textnormal{exp}\left( 5(r_s - r_k)
\right), \\ \nonumber
y_k(r_s) &=& \nu_k\ \textnormal{log}(r_s) + \frac{a_{1,k} + b_{1,k}r_s + c_{1,k}r_s^2  }{ 1 + r_s^2/5} \ , \quad
z_k(r_s) = r_s \frac{ a_{2,k} + b_{2,k}r_s }{ 1 + c_{2,k} r_s^2 } \quad ,
\end{eqnarray}
which becomes exact for $\theta\to 0$ and $\theta\to\infty$, but is limited to the accuracy of the classical mapping data in between. Further, it does not include the exact Hartree-Fock limit for $r_s\to0$, so that it cannot reasonably be used for $r_s<1$.
For completeness, we mention that a functional form similar to Eq.~(\ref{eq:pdw}) has recently been used by Brown \textit{et al.}~\cite{brown3} for a fit to their RPIMC data \cite{brown}.

\begin{table}
\centering
\begin{tabular}{ l c c c c c c c c  }
$k$ & $a_{1,k}$ & $b_{1,k}$ & $c_{1,k}$ & $a_{2,k}$ & $b_{2,k}$ & $c_{2,k}$ & $\nu_k$ & $r_k$ \\ \hline
1 &
5.6304 &
-2.2308 &
1.7624 & 
2.6083 &
1.2782 &
0.16625 &
1.5 &
4.4467 \\
2 &
5.2901 &
-2.0512 &
1.6185 &
-15.076 &
24.929 &
2.0261 &
3 &
4.5581 \\
3 &
3.6854 &
-1.5385 &
1.2629 &
2.4071 &
0.78293 &
0.095869 &
3 &
4.3909 \\

\end{tabular}
\caption{Fit parameters by Perrot and Dharma-wardana \cite{pdw} for the $f_{xc}(r_s,\theta)$ parametrization from Eq.~(\ref{eq:pdw}).}
\label{tab:pdw}
\end{table}
Similar ideas of quantum-classical mappings were recently investigated by Dufty and Dutta, see e.g. \cite{dufty_dutta1, dufty_dutta2}.

\subsection{\label{sub:ksdt}Parametrization by Karasiev \textit{et al.}}
Karasiev and co-workers \cite{karasiev} (KSDT) utilized as the functional form for $f_{xc}$ an expression similar to Eq.~(\ref{eq:ichi_V}), which Ichimaru and co-workers \cite{ichimaru_rev_2,ichimaru_rev} suggested for the interaction energy:
\begin{eqnarray}\label{eq:ksdt}
f_{xc}(r_s,\theta) &=& - \frac{1}{r_s} \frac{ a_\textnormal{HF}(\theta) + b(\theta) r_s^{1/2} + c(\theta) r_s }{ 1 + d(\theta)r_s^{1/2} + e(\theta)r_s }, \\ \nonumber
b(\theta) &=& \textnormal{tanh}\left( \theta^{-1/2} \right) \frac{ b_1 + b_2\theta^2+b_3\theta^4}{ 1 + b_4\theta^2 + \sqrt{1.5}\lambda^{-1}b_3\theta^4 } \ , \quad 
c(\theta) = \left[ c_1 + c_2\ \textnormal{exp}\left( -\frac{c_3}{\theta} \right)\right] e(\theta), \\ \nonumber 
d(\theta) &=& \textnormal{tanh}\left( \theta^{-1/2} \right) \frac{ d_1 + d_2\theta^2+d_3\theta^4}{ 1 + d_4\theta^2 + d_5\theta^4 } \ , \quad 
e(\theta) = \textnormal{tanh}\left( \theta^{-1} \right) \frac{ e_1 + e_2\theta^2+e_3\theta^4}{ 1 + e_4\theta^2 + e_5\theta^4 } \ . \quad 
\end{eqnarray}
Further, instead of fitting to the interaction energy $V$, they used the relation
\begin{equation}\label{eq:exc}
E_{xc}(r_s,\theta) = f_{xc}(r_s,\theta) -\left. \theta \frac{\partial f_{xc}(r_s,\theta)}{\partial\theta}\right|_{r_s} 
\end{equation}
and fitted the rhs.~of Eq.~(\ref{eq:exc}) to the recently published RPIMC data for the exchange correlation energy $E_{xc}$ by Brown \textit{et al.}~\cite{brown} that are available for the parameters $r_s=1,\dots,40$ and $\theta=0.0625,\dots,8$.

\begin{table}
\centering
\begin{tabular}{ c c c c c c c  }

  $b_1$ & $b_2$ & $b_3$ & $b_4$ & $c_1$ & $c_2$ & $c_3$   \\
    \hline
  0.283997 & 
  48.932154 &
  0.370919 &
  61.095357 &
  0.870089 &
  0.193077 &
  2.414644 \vspace{0.2cm} \\

    $d_1$ & $d_2$ & $d_3$ & $d_4$ & $d_5$ & $e_1$ & $e_2$  \\
           \hline
  0.579824 &
  94.537454 & 
  97.839603 &
  59.939999 &
  24.388037 &
  0.212036 &
  16.731249 \vspace{0.2cm}
  \\
    
     $e_3$ & $e_4$ & $e_5$ &
  & & & \\  \hline
     28.485792 &
  34.028876 &
  17.235515 &
  & & & \\
    
  \end{tabular}
\caption{Fit parameters by Karasiev \textit{et al.}~\cite{karasiev} for the $f_{xc}(r_s,\theta)$ parametrization from Eq.~(\ref{eq:ksdt}).}
\label{tab:KSDT}
\end{table}

\section{Results\label{sec:results}}
In this section we analyze the behavior of the analytical approximations for the exchange-correlation free energies that were summarized above by comparison to our recent simulation results that cover the entire relevant density range for temperatures $\Theta \ge 0.5$. These data
have an unprecedented accuracy of the order of $0.1\%$, for details, see Refs.~\cite{dornheim_prl, dornheim_pop_17}.

\subsection{\label{sub:temperature}Temperature dependence}

 \begin{figure}[t]
 \includegraphics[width=0.5\textwidth]{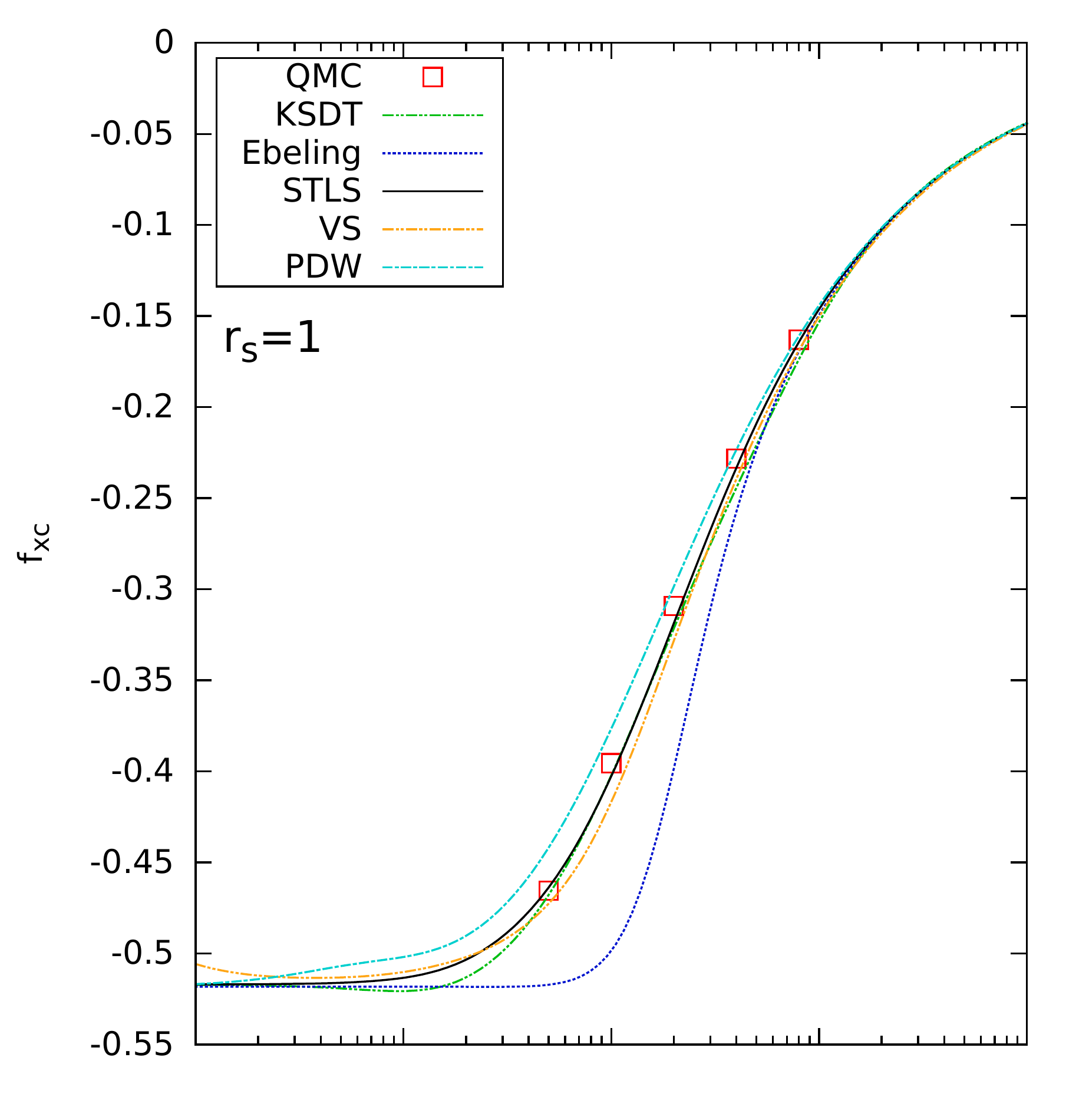} \includegraphics[width=0.5\textwidth]{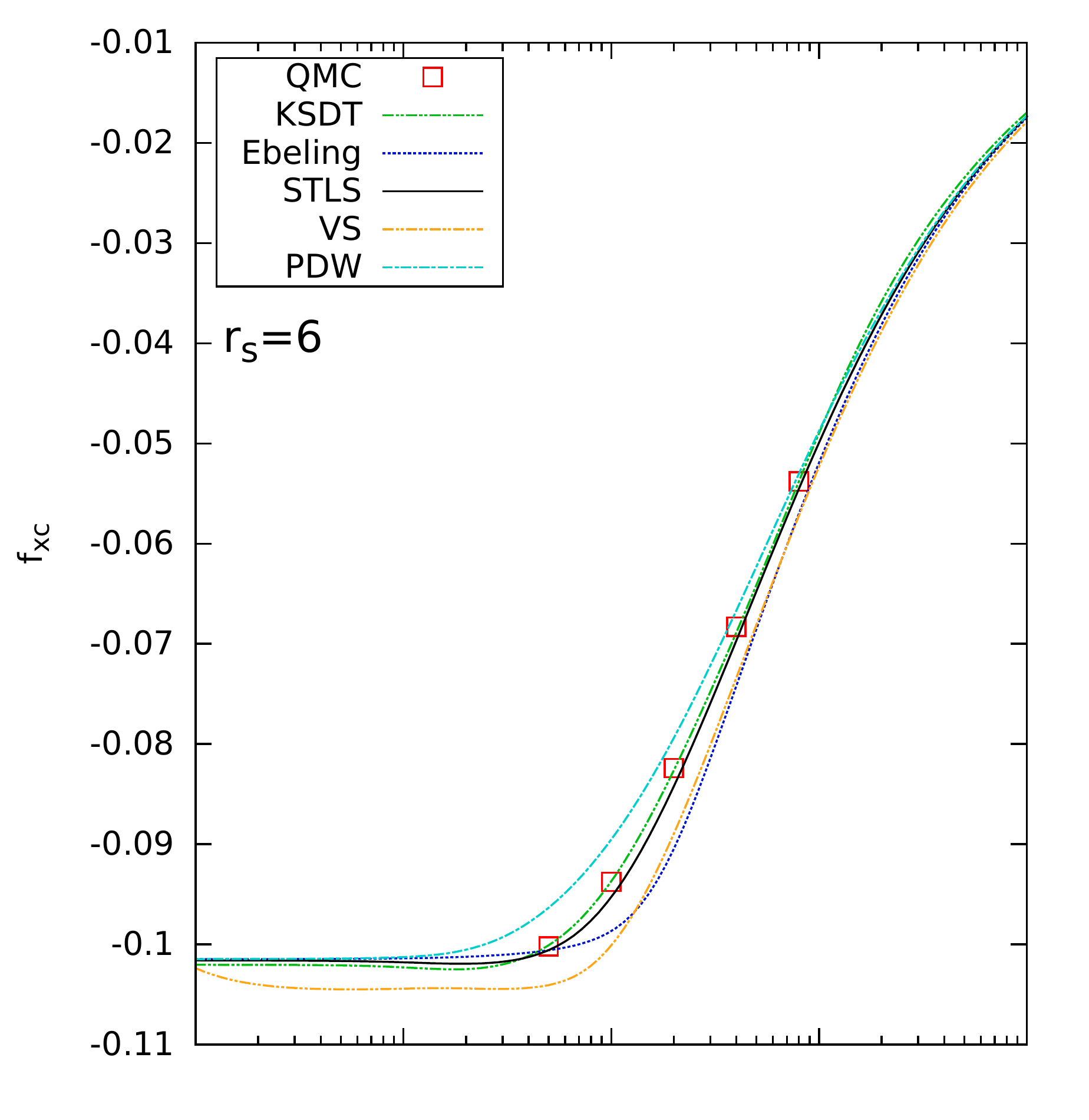}~\hspace{-3mm}\vspace*{-0.6cm}\\ 
  \includegraphics[width=0.5\textwidth]{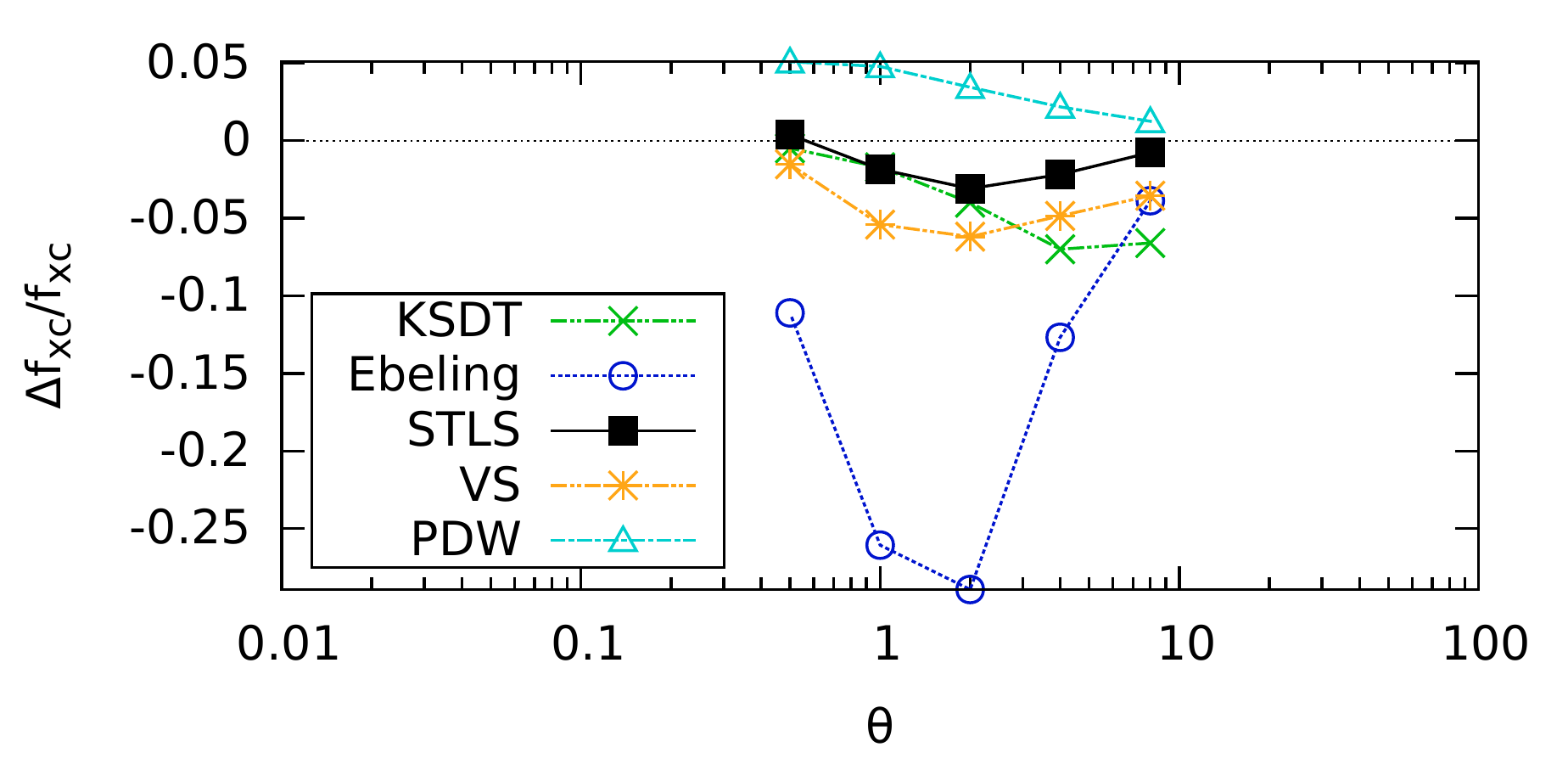}
    \includegraphics[width=0.5\textwidth]{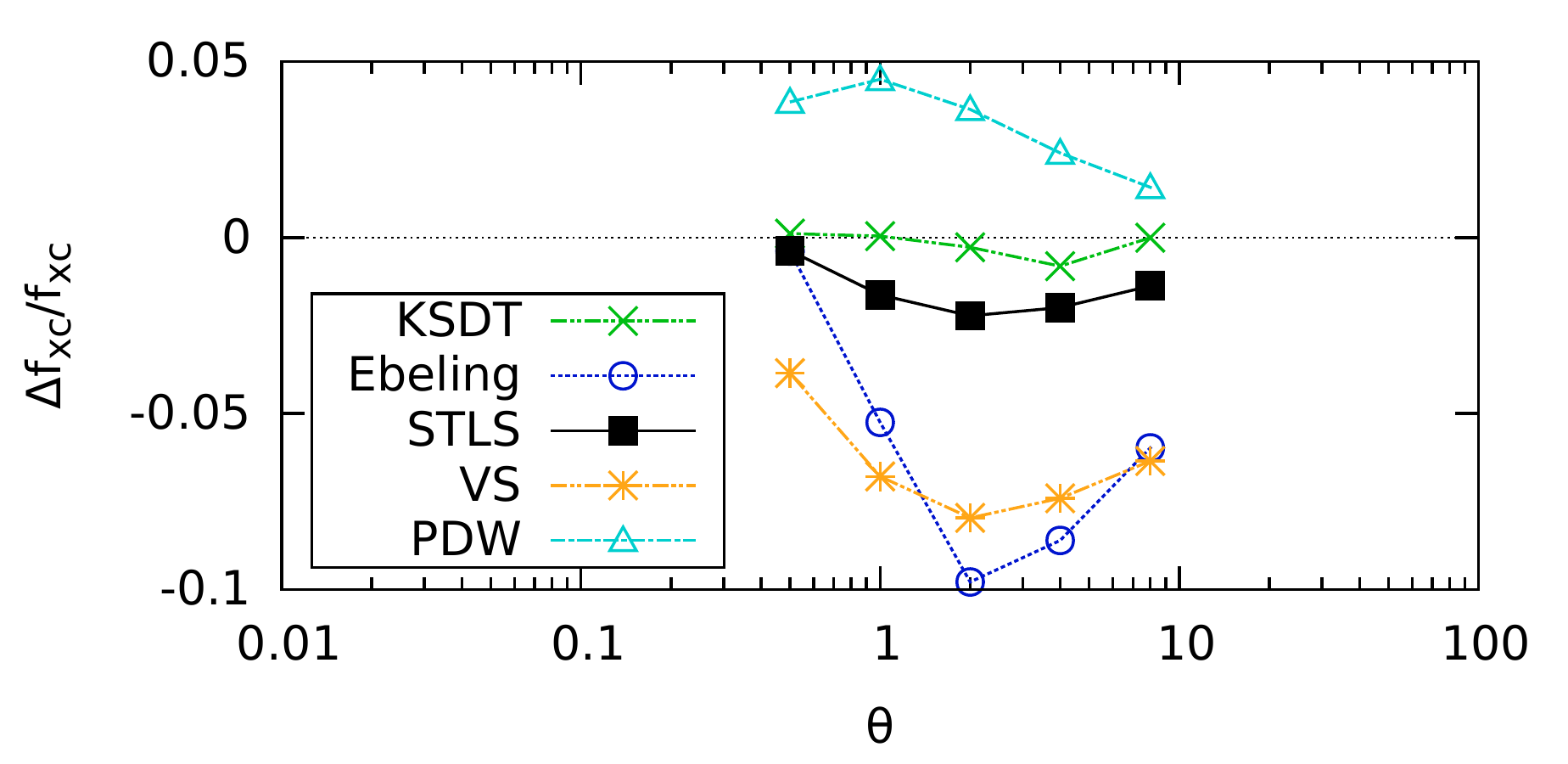}~\hspace{-3mm}
 \caption{Temperature dependence of $f_{xc}$ at fixed density $r_s=1$ (left) and $r_s=6$ (right). Top: QMC data (symbols) taken from Dornheim \textit{et al.}\cite{dornheim_prl}, a parametrization of RPIMC data by Karasiev \textit{et al.}\cite{karasiev} (KSDT), a semi-analytic Pad\'e approximation by Ebeling \cite{ebeling1}, a parametrization fitted to STLS and VS data by Ichimaru \cite{ichimaru_rev} and Sjostrom and Dufty \cite{stls2}, respectively, and a fit to classical mapping data by Perrot and Dharma-wardana \cite{pdw} (PDW).   Bottom: Relative deviation to the QMC data.
  \label{fig:theta} }
 \end{figure}

In Fig.~\ref{fig:theta}, we show the temperature dependence of the  exchange-correlation free energy as a function of the reduced temperature $\theta$ for two densities that are relevant for contemporary warm dense matter research, namely $r_s=1$ (left) and $r_s=6$ (right).
For both cases, all depicted parametrizations reproduce the correct classical limit for large $\theta$  [c.f.~Eq.~(\ref{eq:DBH})] and four of them (Ebeling, KSDT, STLS and PDW) are in excellent agreement for the ground state as well. 
For completeness, we note that the small differences between KSDT and Ebeling and PDW are due to different ground state QMC input data. In particular, Karasiev \textit{et al.}~used more recent QMC results by Spink \textit{et al.}\cite{spink}, although in the context of WDM research the deviations to older parametrizations are negligible. Further, we note that the STLS parametrization reproduces the STLS data for $\theta=0$ that, however, are in good agreement with the exact QMC results as well. The VS-parametrization, on the other hand, does not incorporate any ground state limit and, consequently, the behavior of $f^\textnormal{VS}_\textnormal{xc}(r_s,\theta)$ becomes unreasonable below $\theta=0.0625$.
Similarly, the lowest temperature (despite the ground state limit) included in the fit for $f^\textnormal{PDW}_\textnormal{xc}(r_s,\theta)$ is $\theta=0.25$ and the rather unsmooth connection between this point and $\theta=0$ does not appear to be trustworthy as well.

Let us now check the accuracy of the different models at intermediate, WDM temperatures. As a reference, we use the recent, accurate QMC results for the macroscopic UEG by Dornheim \textit{et al.}~\cite{dornheim_prl}, i.e., the red squares. For $r_s=1$, the semi-analytic expression by Ebeling (blue) exhibits the largest deviations exceeding $\Delta f_{xc}/f_{xc}= 25\%$ for $\theta\sim 1$.
For lower density, $r_s=6$, the Ebeling parametrization is significantly more accurate although here, too, appear deviations of $\Delta f_{xc}/f_{xc}\sim 10\%$ to the exact data at intermediate temperature. Therefore, this parametrization produces reliable data in the two limiting cases of zero and high temperature, but is less accurate in between.

Next consider the STLS curve (black). It is in very good agreement with the QMC data, and the error does not exceed $\Delta f_{xc}/f_{xc} = 4\%$ over the entire $\theta$-range for both depicted $r_s$ values. The largest deviations appear for intermediate temperatures as well.

Third, we consider the VS-model (yellow line). For $r_s=1$, the VS-parametrization  by Sjostrom and Dufty \cite{stls2} exhibits the same trends as the STLS curve albeit with larger deviations, $\Delta f_{xc}/f_{xc}> 5\%$. Further, for $r_s=6$, $f_{xc}^\textnormal{VS}$ exhibits much larger deviations to the exact result and the error attains $\Delta f_{xc}/f_{xc}\approx 8\%$. Evidently, the constraint to automatically fulfill the CSR does not improve the accuracy of other quantities, in particular the interaction energy $V$ (which was used as an input for the parametrization, see Sec.~\ref{sub:vs}) or the static structure factor $S(k)$ itself.   

Fourth, the parametrization based on the classical mapping (PDW, light blue) exhibits somewhat  opposite trends as compared to Ebeling, STLS, and VS and predicts a too large exchange correlation free energy for all $\theta$.
The magnitude of the deviations is comparable to VS and does not exceed $\Delta f_{xc}/f_{xc}=5\%$.

Finally, we consider the recent parametrization by Karasiev \textit{et al.}~(KSDT, green) \cite{karasiev} that is based on RPIMC results \cite{brown}.
For $r_s=6$, there is excellent agreement with the new reference QMC data with a maximum deviation of $\Delta f_{xc}/f_{xc}\sim 1\%$ for $\theta=4$. This is, in principle, expected since the main source of error for their input data, i.e., the nodal error and the insufficient finite-size correction, are less important for larger $r_s$. However, for $r_s=1$ there appear significantly larger deviations exceeding $\Delta f_{xc}/f_{xc}= 5\%$ at high temperature. In fact, for $r_s=1$ and the largest considered temperature, $\theta=8$, the KSDT parametrization exhibits the largest deviations of all depicted parametrizations.

\subsection{\label{sub:density}Density dependence}

 \begin{figure}[t]
 \includegraphics[width=0.5\textwidth]{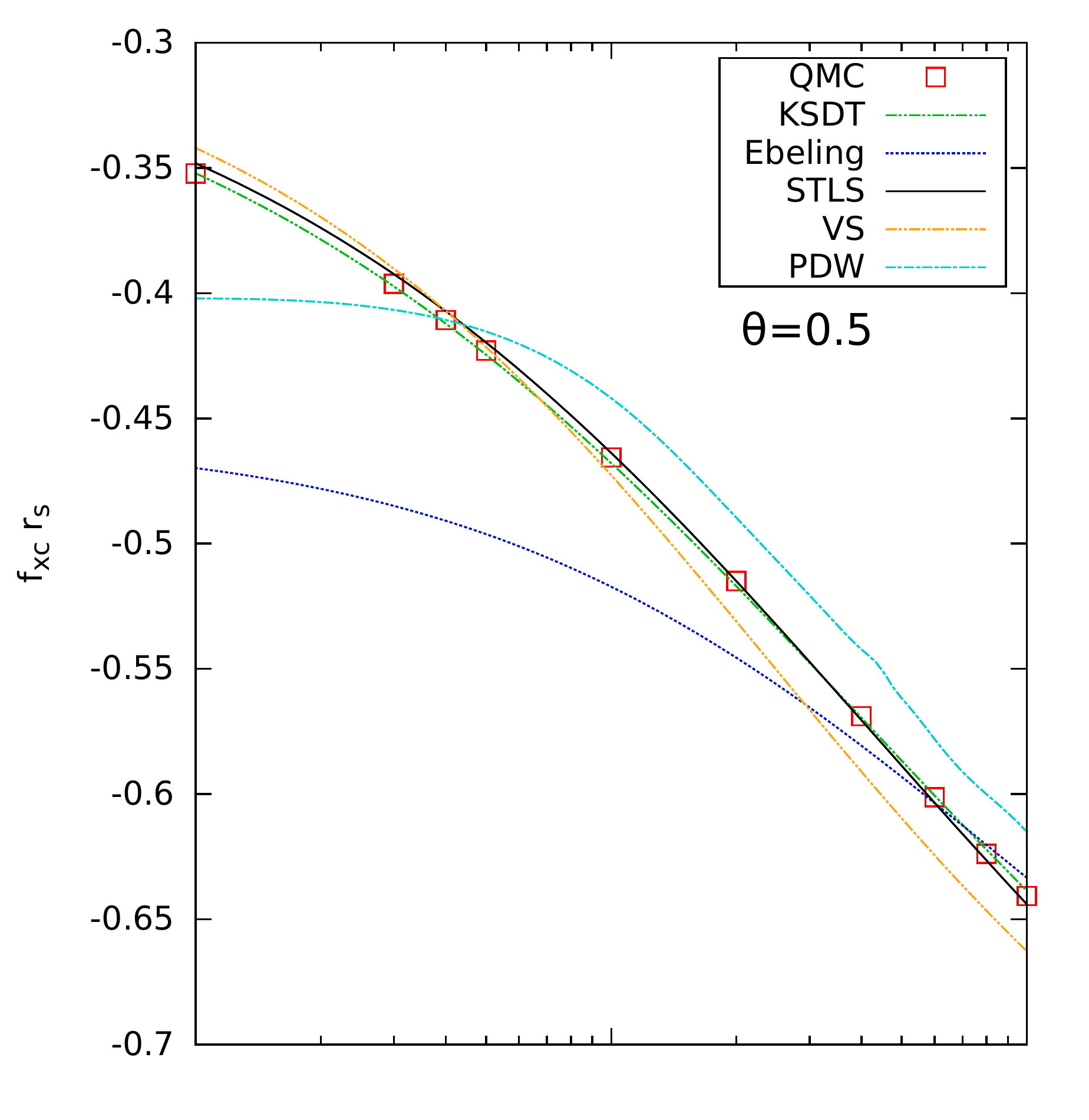} \includegraphics[width=0.5\textwidth]{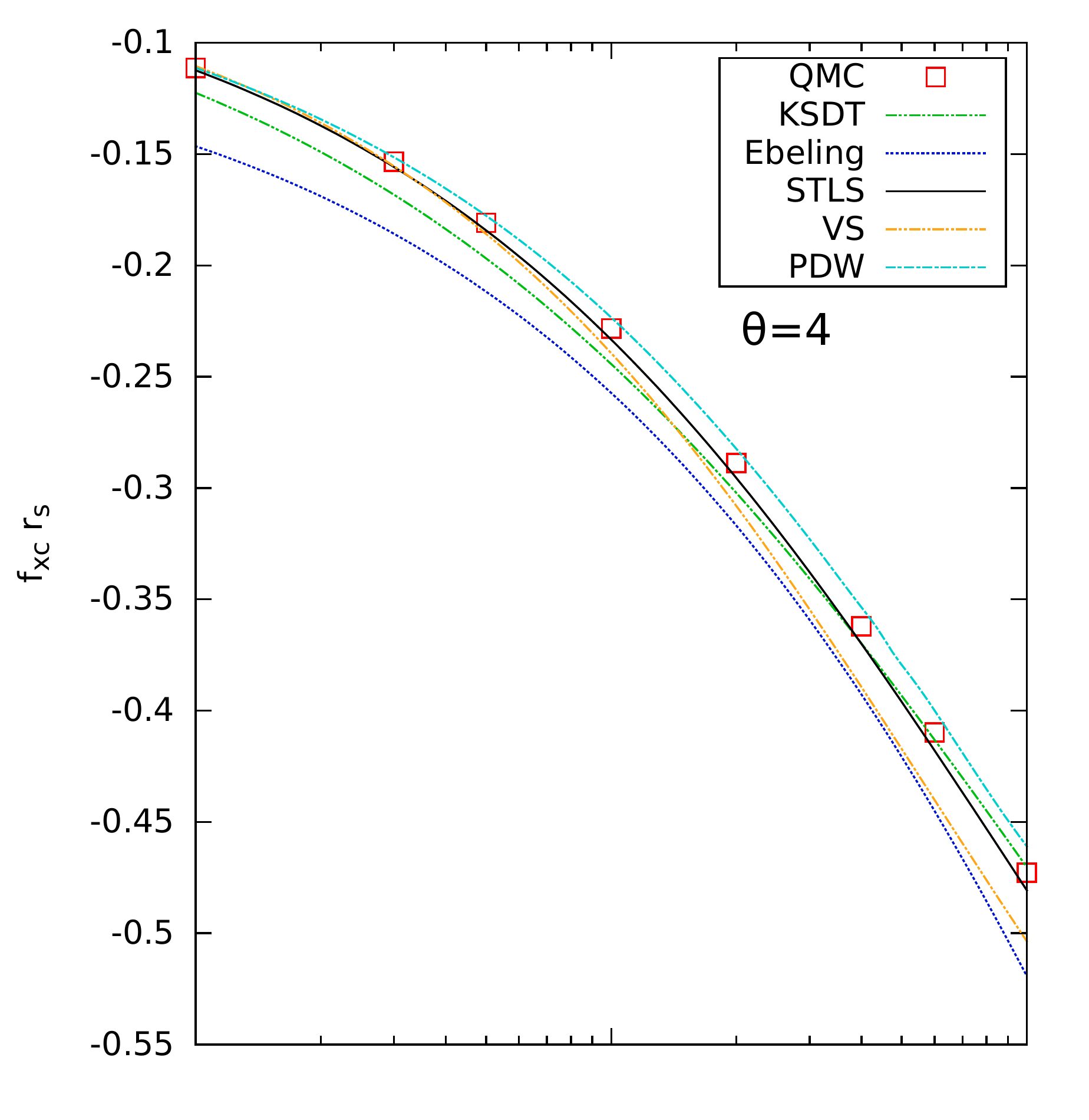}~\hspace{-3mm}\vspace*{-0.6cm}\\ 
  \includegraphics[width=0.5\textwidth]{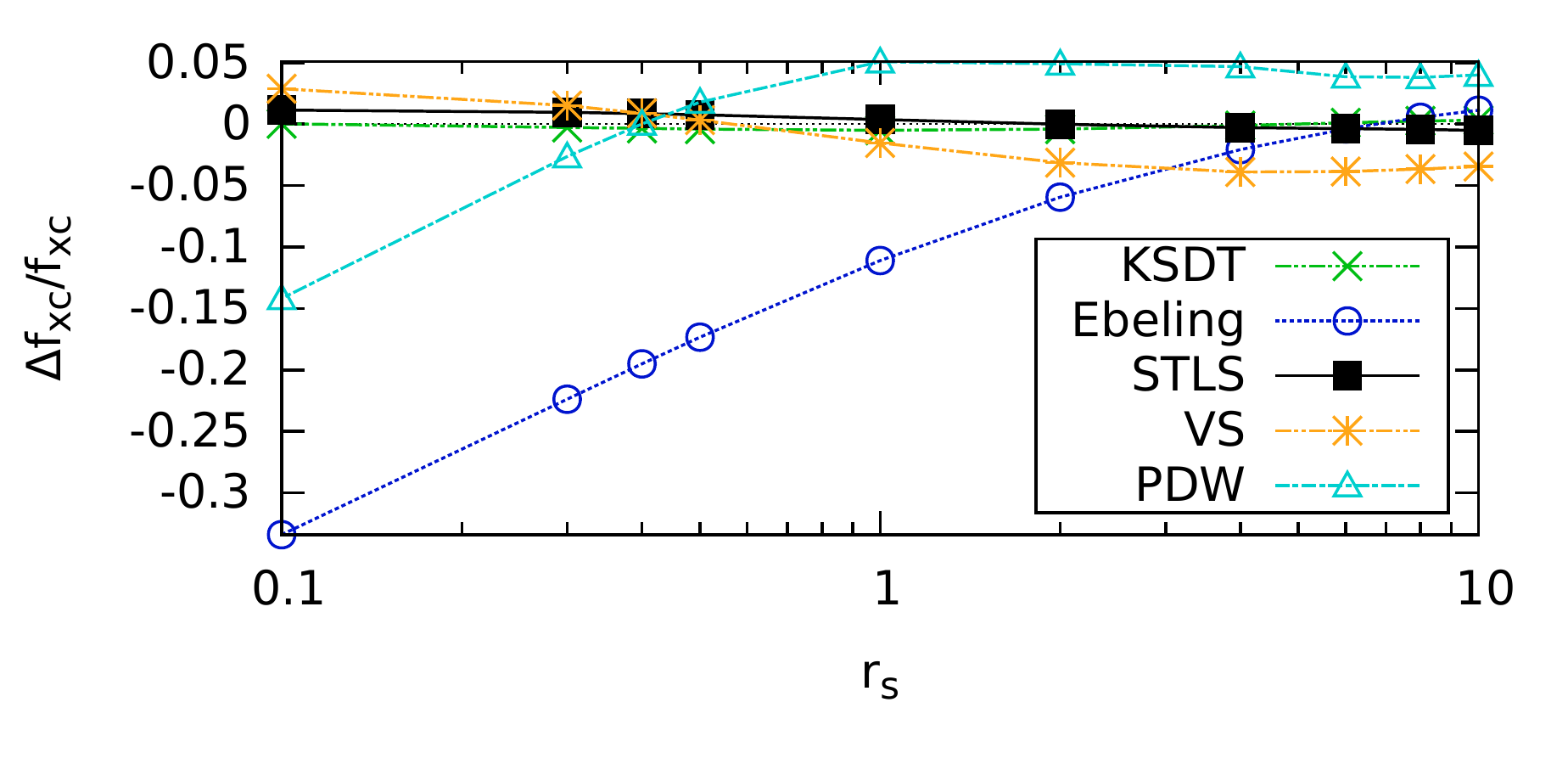}
    \includegraphics[width=0.5\textwidth]{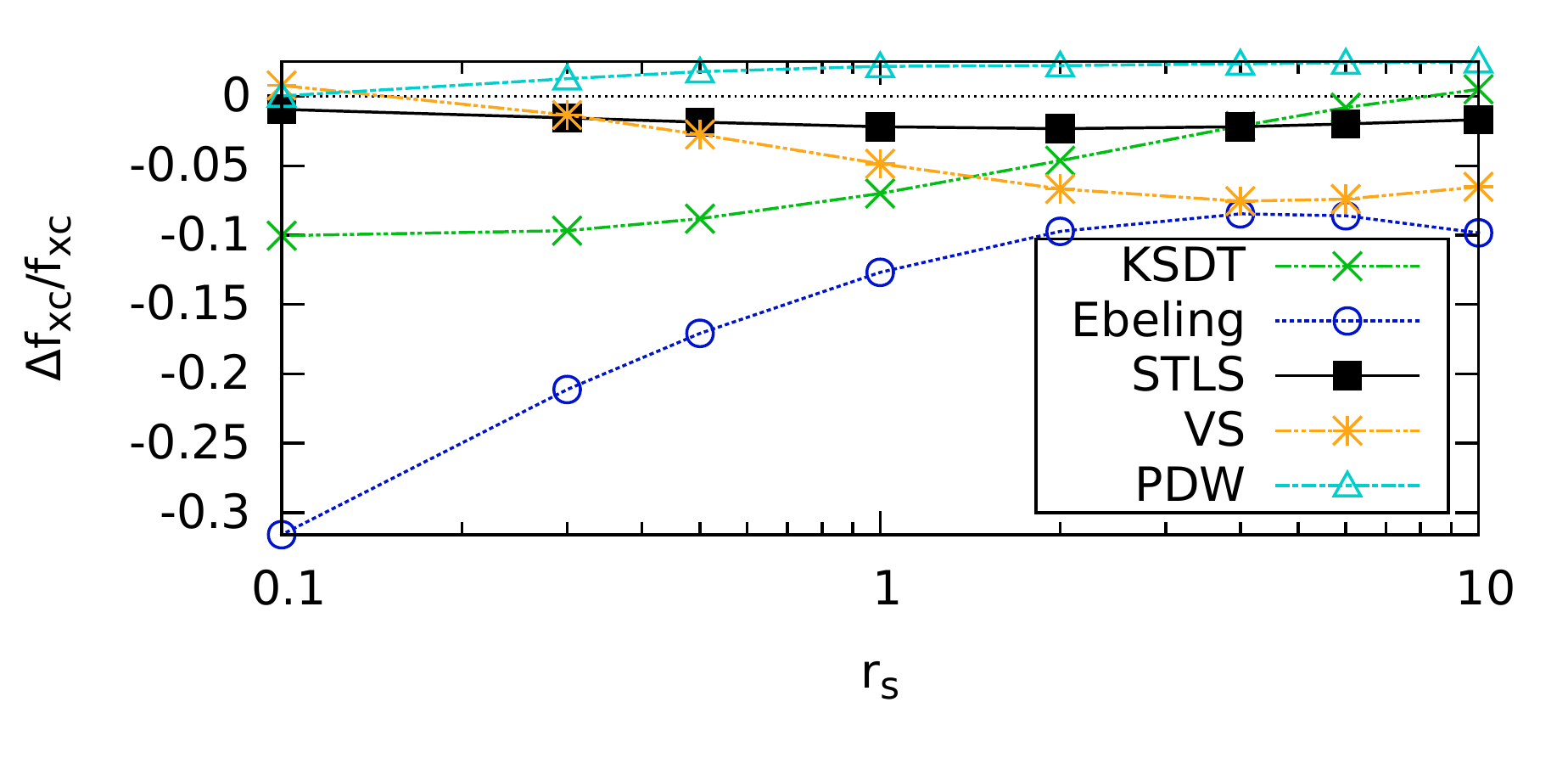}~\hspace{-3mm}
 \caption{Density dependence of $f_{xc}$ at fixed temperature $\theta=0.5$ (left) and $\theta=4$ (right). Top: QMC data taken from Dornheim \textit{et al.}\cite{dornheim_prl}, a parametrization of RPIMC data by Karasiev \textit{et al.}\cite{karasiev} (KSDT), a semi-analytic Pad\'e approximation by Ebeling \cite{ebeling1}, a parametrization fitted to STLS and VS data by Ichimaru \cite{ichimaru_rev} and Sjostrom and Dufty \cite{stls2}, respectively, and a fit to classical mapping data by Perrot and Dharma-wardana \cite{pdw} (PDW).   Bottom: Relative deviation to the QMC data.
  \label{fig:density}}
 \end{figure}

As a complement to Sec.~\ref{sub:temperature}, in Fig.~\ref{fig:density} we investigate in more detail the density dependence of the different parametrizations for two relevant temperatures, $\theta=0.5$ (left) and $\theta=4$ (right).

Most notably the Ebeling and PDW parametrizations do not include the correct high density ($r_s\to0$) limit, i.e.~Eq.~(\ref{eq:ichi_definitions}), and therefore are not reliable for $r_s<1$. 
For $\theta=0.5$, $f_{xc}^\textnormal{Ebeling}$ is in qualitative agreement with the correct results, but the deviations rapidly increase with density and exceed $\Delta f_{xc}/f_{xc} = 10\%$, for $r_s=1$. At higher temperature, $\theta=4$, the situation is worse, and the Ebeling parametrization shows systematic deviations over the entire density range.
The STLS fit displays a similarly impressive agreement with the exact data as for the $\theta$-dependence (c.f.~Fig.~\ref{fig:theta}), and the deviations do not exceed $\Delta f_{xc}/f_{xc}\sim 3\%$ for both depicted $\theta$-values.
On the other hand, the VS results are again significantly less accurate than STLS although the deviation remains below $\Delta f_{xc}/f_{xc}=8\%$ for both temperatures. Further, we notice that the largest deviations occur for $r_s \geq 2$, i.e., towards stronger coupling, which is expected since here the pair distribution function exhibits unphysical negative values at short distance, e.g.~\cite{stls2}.
Again, the incorporation of the CSR has not improved the quality of the interaction energy or the structure factor compared to STLS.
The classical mapping data (PDW) does exhibit deviations not exceeding $\Delta f_{xc}/f_{xc} =5\%$ for $r_s\geq1$, i.e., in the range where numerical data have been incorporated into the fit. Overall, the quality of this parametrization is comparable to the VS-curve although the relative deviation appears to be almost constant with respect to the density. This is not surprising as the approximation has not been conducted with respect to coupling (the effective classical system is solved with the hypernetted chain method, which is expected to be accurate in this regime) but, instead, in the interpolation of the effective temperature $T_c$. Further, we notice a peculiar non-smooth and almost oscillatory behavior of $f^\textnormal{PDW}_{xc}$ around $r_s=5$ that is more pronounced for $\theta=0.5$ and the origin of which remains unclear. 
Finally, we again consider the KSDT-fit based on the RPIMC data by Brown \textit{et al.}~\cite{brown} (a similar analysis for more temperatures can be found in Ref.~\cite{dornheim_prl}).
For $\theta=0.5$, this parametrization is in excellent agreement with the reference QMC data and the deviations are in the sub-percent regime over the entire depicted $r_s$-range.
However, for larger temperatures there appear significant errors that, at $\theta=4$, attain a maximum of $\Delta f_{xc}/f_{xc}\sim 10\%$ for $r_s=0.1$, i.e., at parameters where STLS, VS, and PDW are in very good agreement with the reference QMC data. Interestingly, these deviations only vanish for $r_s \leq 10^{-4}$.  
Naturally, the inaccuracies of the KSDT-fit are a direct consequence of the systematic errors of the input data and the lack of accurate simulation data for $r_s<1$, prior to Ref.~\cite{dornheim_prl}.

\section{Discussion}

In summary, we have compared five different parametrizations of the  exchange-correlation free energy of the unpolarized UEG to the recent QMC data by Dornheim \textit{et al.}~\cite{dornheim_prl} and, thereby, have been able to gauge their accuracy with respect to $\theta$ and $r_s$ over large parts of the warm dense matter regime. We underline that all these parametrizations are highly valuable, the main merit being their easy and flexible use and rapid evaluation. At the same time, an unbiased evaluation of their accuracy had not been done and appears highly important, as this allows to constrain the field of applicability of these models and to indicate directions for future improvements.

Summarizing our findings, 
we have observed that the semi-analytic parametrization by Ebeling \cite{ebeling1} is mostly reliable in the high- and zero temperature limits, but exhibits substantial deviations in between.
The STLS-fit given by Ichimaru an co-workers \cite{ichimaru_rev_2,ichimaru_rev}, on the other hand, exhibits a surprisingly high accuracy for all investigated $r_s$-$\theta$-combinations with a typical relative systematic error of $\sim2\%$. The more recent Vashishta-Singwi (VS) results \cite{stls2} that automatically fulfill the compressibility sum-rule display a qualitatively similar behavior but are significantly less accurate everywhere.
The classical mapping suggested by Perrot and Dharma-wardana \cite{pdw} constitutes an approximation rather with respect to temperature than to coupling strength and, consequently, exhibits different trends. In particular, we have found that the relative systematic error is nearly independent of $r_s$, but decreases with increasing $\theta$ and eventually vanishes for $\theta\to\infty$. Overall, the accuracy of the PDW-parametrization is comparable to VS and, hence, inferior to STLS. Finally, the more recent fit by Karasiev \textit{et al.}~\cite{karasiev} to RPIMC data \cite{brown} is accurate for large $r_s$ and low temperature, where the input data is not too biased by the inappropriate treatment of finite-size errors in the underlying RPIMC results. 
For higher temperatures (where the exchange correlation free energy constitutes only a small fraction of the total free energy) there occur relative deviations of up to $\sim10\%$.

Thus we conclude that an accurate parametrization of the exchange-correlation free energy that is valid for all $r_s$-$\theta$-combinations is presently not available. However, the recent QMC data by Dornheim \textit{et al.}~\cite{dornheim_prl} most certainly constitute a promising basis for the construction of such a functional. Further, thermal DFT calculations in the local spin-density approximation require a parametrization of $f_{xc}$ also as a function of the spin-polarization $\xi = (N_\uparrow - N_\downarrow) / (N_\uparrow + N_\downarrow)$, i.e., $f_{xc}(r_s,\theta,\xi)$ for all warm dense matter parameters. Obviously, this will require to extend the QMC simulations beyond the unpolarized case, $\xi\in(0,1]$ and, in addition, reliable data for $\theta<0.5$ are indispensable. This work is presently under way.
We also note that the quality of the currently available KSDT fit for $f_{xc}(r_s,\theta,\xi)$ remains to be tested for $\xi> 0$. The accuracy of this parametrization is limited by (i) the quality of the RPIMC data (for the spin-polarized UEG ($\xi=1$) they are afflicted with a substantially larger nodal error than for the unpolarized case that we considered in the present paper, see Ref.~\cite{groth}) and (ii) by the quality of the PDW-results \cite{pdw} that have been included as the only input to the KSDT-fit for $0<\xi<1$ at finite $\theta$.
Therefore, we conclude that the construction of a new, accurate function $f_{xc}(r_s,\theta,\xi)$ is still of high importance for thermal DFT and semi-analytical models, for comparisons with experiments, but also for explicitly time-dependent approaches such as time-dependent DFT and quantum hydrodynamics \cite{manfredi, michta}.

\begin{acknowledgement}
SG and TD contributed equally to this work.
 We acknowledge helpful comments from A. F\"orster on the Pad\'e formulas of Ebeling \textit{et al.}~and from Fionn~D.~Malone.
 This work was supported by the Deutsche Forschungsgemeinschaft via project BO1366-10 and via SFB TR-24 project A9 as well as grant shp00015 for CPU time at the Norddeutscher Verbund f\"ur Hoch- und H\"ochstleistungsrechnen (HLRN).
\end{acknowledgement}


\begin{thebibliography}{26}













\bibitem{knudson} M.D.~Knudson \textit{et al.}, Probing the Interiors of the Ice Giants: Shock Compression of Water to 700 GPa and $3.8\mathbf{g}/{{cm}}^{3}$, \href{ http://link.aps.org/doi/10.1103/PhysRevLett.108.091102 }{ \textit{Phys. Rev. Lett.} \textbf{108}, 091102} (2012)

\bibitem{militzer} B.~Militzer \textit{et al.}, A Massive Core in Jupiter Predicted from First-Principles Simulations, \href{ http://iopscience.iop.org/1538-4357/688/1/L45 }{\textit{Astrophys. J.} \textbf{688}, L45} (2008)





\bibitem{ernst} R.~Ernstorfer \textit{et al.}, The Formation of Warm Dense Matter: Experimental Evidence for Electronic Bond Hardening in Gold, \href{ http://www.sciencemag.org/content/323/5917/1033}{ \textit{Science} \textbf{323}, 5917} (2009)




\bibitem{nora} R.~Nora \textit{et al.}, Gigabar Spherical Shock Generation on the OMEGA Laser \href{ http://link.aps.org/doi/10.1103/PhysRevLett.114.045001 }{ \textit{Phys. Rev. Lett.} \textbf{114}, 045001} (2015)


\bibitem{schmit} P.F.~Schmit \textit{et al.}, Understanding Fuel Magnetization and Mix Using Secondary Nuclear Reactions in Magneto-Inertial Fusion \href{ http://link.aps.org/doi/10.1103/PhysRevLett.113.155004 }{ \textit{Phys. Rev. Lett.} \textbf{113}, 155004} (2014)



\bibitem{hurricane3} O.A.~Hurricane \textit{et al.}, Inertially confined fusion plasmas dominated by alpha-particle self-heating, \href{ http://www.nature.com/nphys/journal/vaop/ncurrent/full/nphys3720.html }{ \textit{Nature Phys.}~3720 } (2016) 






\bibitem{ks} W.~Kohn, and L.J.~Sham, Self-Consistent Equations Including Exchange and Correlation Effects, \href{http://link.aps.org/doi/10.1103/PhysRev.140.A1133}{\textit{Phys.Rev.}~\textbf{140}, A1144} (1965)

\bibitem{dft_review} R.O.~Jones, Density functional theory: Its origins, rise to prominence, and future, \href{ http://link.aps.org/doi/10.1103/RevModPhys.87.897}{ \textit{Rev.~Mod.~Phys.}~\textbf{87}, 897-923 } (2015)



\bibitem{alder} D.M.~Ceperley and B.J.~Alder, Ground State of the Electron Gas by a Stochastic Method, \href{ http://link.aps.org/doi/10.1103/PhysRevLett.45.566}{ \textit{Phys. Rev. Lett.} \textbf{45}, 566} (1980)


\bibitem{perdew} J.P.~Perdew and A.~Zunger, Self-interaction correction to density-functional approximations for many-electron systems, \href{ http://link.aps.org/doi/10.1103/PhysRevB.23.5048 }{ \textit{Phys.~Rev.~B} \textbf{23}, 5048 } (1981)






\bibitem{mermin} N.D.~Mermin, Thermal Properties of the Inhomogeneous Electron Gas, \href{http://link.aps.org/doi/10.1103/PhysRev.137.A1441}{\textit{Phys. Rev.}~\textbf{137}, A1441} (1965)



















\bibitem{karasiev2} V.V.~Karasiev, L.~Calderin, and S.B.~Trickey, The importance of finite-temperature exchange-correlation for warm dense matter calculations, \href{ http://link.aps.org/doi/10.1103/PhysRevE.93.063207 }{ \textit{Phys.~Rev.~E} \textbf{93}, 063207 } (2016)



\bibitem{dharma} M.W.C.~Dharma-wardana, Current Issues in Finite-T Density-Functional Theory and Warm-Correlated Matter, \href{ http://www.mdpi.com/2079-3197/4/2/16 }{ \textit{Computation} \textbf{4}, 16} (2016)

\bibitem{dharma_cpp15} M. W. C. Dharma~wardana,
A Review of Studies on Strongly-Coupled Coulomb Systems Since the Rise of DFT and SCCS-1977, 
\href{http://onlinelibrary.wiley.com/doi/10.1002/ctpp.201400073/abstract }{ \textit{Contrib.~Plasma Phys.}~{\bf 55}, 85-101} (2015)


\bibitem{gga} T.~Sjostrom and J.~Daligault, Gradient corrections to the exchange-correlation free energy, \href{ http://link.aps.org/doi/10.1103/PhysRevB.90.155109 }{ \textit{Phys.~Rev.~B} \textbf{90}, 155109 } (2014)


\bibitem{burke} K.~Burke, J.C.~Smith, P.E.~Grabowski, and A.~Pribram-Jones, Exact conditions on the temperature dependence of density functionals, \href{ http://link.aps.org/doi/10.1103/PhysRevB.93.195132 }{ \textit{Phys.~Rev.~B} \textbf{93}, 195132 } (2016)



\bibitem{burke2} A.~Pribram-Jones, P.E.~Grabowski, and K.~Burke, Thermal Density Functional Theory: Time-Dependent Linear Response and Approximate Functionals from the Fluctuation-Dissipation Theorem, \href{ http://link.aps.org/doi/10.1103/PhysRevLett.116.233001 }{ \textit{Phys.~Rev.~Lett.}~\textbf{116}, 233001} (2016)












\bibitem{tim2} T.~Schoof, S.~Groth and M.~Bonitz, Towards ab Initio Thermodynamics of the Electron Gas at Strong Degeneracy, \href{http://onlinelibrary.wiley.com/doi/10.1002/ctpp.201400072/abstract}{ \textit{Contrib. Plasma Phys.} \textbf{55}, 136-143} (2015)

\bibitem{tim_prl} T.~Schoof, S.~Groth, J.~Vorberger and M.~Bonitz, \textit{Ab Initio} Thermodynamic Results for the Degenerate Electron Gas at Finite Temperature, \href{ http://link.aps.org/doi/10.1103/PhysRevLett.115.130402}{ \textit{Phys.~Rev.~Lett.}~\textbf{115}, 130402} (2015)



\bibitem{dornheim} T.~Dornheim, S.~Groth, A.~Filinov and M.~Bonitz, Permutation blocking path integral Monte Carlo: a highly efficient approach to the simulation of strongly degenerate non-ideal fermions, \href{ http://iopscience.iop.org/1367-2630/17/7/073017 }{ \textit{New J. Phys.} \textbf{17}, 073017} (2015)


\bibitem{dornheim2} T.~Dornheim, T.~Schoof, S.~Groth, A.~Filinov, and M.~Bonitz, Permutation Blocking Path Integral Monte Carlo Approach to the Uniform Electron Gas at Finite Temperature, \href{ http://scitation.aip.org/content/aip/journal/jcp/143/20/10.1063/1.4936145 }{ \textit{ J. Chem. Phys.} \textbf{143}, 204101} (2015)



\bibitem{groth} S.~Groth, T.~Schoof, T.~Dornheim, and M.~Bonitz, \textit{Ab Initio} Quantum Monte Carlo Simulations of the Uniform Electron Gas without Fixed Nodes, \href{ http://link.aps.org/doi/10.1103/PhysRevB.93.085102 }{ \textit{Phys.~Rev.~B} \textbf{93}, 085102} (2016)




\bibitem{dornheim3} T.~Dornheim, S.~Groth, T.~Schoof, C.~Hann, and M.~Bonitz, \textit{Ab initio} quantum Monte Carlo simulations of the Uniform electron gas without fixed nodes: The unpolarized case, \href{ http://link.aps.org/doi/10.1103/PhysRevB.93.205134 }{ \textit{Phys.~Rev.~B} \textbf{93}, 205134 } (2016)





\bibitem{dornheim_cpp} T.~Dornheim, H.~Thomsen, P.~Ludwig, A.~Filinov, and M.~Bonitz, Analyzing Quantum Correlations made simple, \href{ http://onlinelibrary.wiley.com/doi/10.1002/ctpp.201500120/abstract }{ \textit{Contrib.~Plasma~Phys.}~\textbf{56}, 371} (2016)





\bibitem{malone} F.D.~Malone \textit{et al.}, Interaction Picture Density Matrix Quantum Monte Carlo, \href{ http://scitation.aip.org/content/aip/journal/jcp/143/4/10.1063/1.4927434 }{ \textit{J.~Chem.~Phys.}~\textbf{143}, 044116 } (2015)

\bibitem{malone2} Fionn~D.~Malone, N.S.~Blunt, Ethan~W.~Brown, D.K.K.~Lee, J.S.~Spencer, W.M.C.~Foulkes, and James J.~Shepherd, Accurate Exchange-Correlation Energies for the Warm Dense Electron Gas, \href{http://link.aps.org/doi/10.1103/PhysRevLett.117.115701}{\textit{Phys.~Rev.~Lett.} \textbf{117}, 115701 } (2016)








\bibitem{dornheim_prl} T.~Dornheim, S.~Groth, T.~Sjostrom, F.D.~Malone, W.M.C.~Foulkes, and M.~Bonitz,  \textit{Ab Initio} Quantum Monte Carlo Simulation of the Warm Dense Electron Gas in the Thermodynamic Limit,  \href{ http://link.aps.org/doi/10.1103/PhysRevLett.117.156403 }{\textit{Phys.~Rev.~Lett.}~\textbf{117}, 156403} (2016)


\bibitem{dornheim_pop_17}  T.~Dornheim, S.~Groth,  F.D.~Malone, T.~Schoof, T.~Sjostrom, W.M.C.~Foulkes, and M.~Bonitz,  \textit{Ab Initio} Quantum Monte Carlo Simulation of the Warm Dense Electron Gas, submitted to Phys. Plasmas, arXiv:1611.02658. 















\bibitem{loh} E.Y.~Loh, J.E.~Gubernatis, R.T.~Scalettar, S.R.~White, D.J.~Scalapino and R.L.~Sugar, Sign problem in the numerical simulation of many-electron systems, \href{http://link.aps.org/doi/10.1103/PhysRevB.41.9301}{\textit{Phys. Rev. B} \textbf{41}, 9301-9307} (1990)

\bibitem{troyer} M.~Troyer and U.J.~Wiese, Computational Complexity and Fundamental Limitations to Fermionic Quantum Monte Carlo Simulations, \href{http://link.aps.org/doi/10.1103/PhysRevLett.94.170201}{\textit{Phys. Rev. Lett.} \textbf{94}, 170201} (2005)






\bibitem{kraeft-book} W.-D.~Kraeft, D.~Kremp, W.~Ebeling, and G.~R\"opke, Quantum Statistics of Charged Particle Systems, Akademie-Verlag Berlin (1986)



\bibitem{dewitt-tribute} Heather D. Whitley, Angel Alastuey, Jim A. Gaffney, Robert Cauble, Wolf-Dietrich Kraeft, and Michael Bonitz,
A tribute to pioneers of strongly coupled plasmas: Hugh E. DeWitt, Bernard Jancovici, and Forrest J. Rogers
\href{ http://onlinelibrary.wiley.com/doi/10.1002/ctpp.201400083/abstract }{ \textit{Contrib.~Plasma Phys.}~{\bf 55}, 102-115} (2015)




\bibitem{ebeling1} W.~Ebeling, Nonideality Effects in Plasmas with Multiply Charged Ions, \href{ http://onlinelibrary.wiley.com/doi/10.1002/ctpp.2150290206/abstract}{ \textit{Contrib.~Plasma Phys.}~\textbf{29}, 165} (1989)





\bibitem{stls_original} K.S.~Singwi, M.P.~Tosi, R.H.~Land, and A.~Sj\"olander, Electron Correlations at Metallic Densities, \href{ http://link.aps.org/doi/10.1103/PhysRev.176.589 }{ \textit{Phys.~Rev.~}\textbf{176}, 589-599 } (1968)





\bibitem{vs_original} P.~Vashishta and K.S.~Singwi, Electron Correlations at Metallic Densities. V, \href{ http://link.aps.org/doi/10.1103/PhysRevB.6.875 }{ \textit{Phys.~Rev.~B} \textbf{6}, 875} (1972)






\bibitem{pdw_prl} M.W.C.~Dharma-wardana and F.~Perrot, Simple Classical Mapping of the Spin-Polarized Quantum Electron Gas: Distribution Functions and Local-Field Corrections, \href{ http://link.aps.org/doi/10.1103/PhysRevLett.84.959 }{ \textit{Phys.~Rev.~Lett.}~\textbf{84}, 959 } (2000)


\bibitem{pdw} F.~Perrot, M.W.C.~Dharma-wardana, Spin-polarized electron liquid at arbitrary temperatures:  Exchange-correlation energies, electron-distribution functions, and the static response functions, \href{http://link.aps.org/doi/10.1103/PhysRevB.62.16536}{\textit{Phys.~Rev.~B} \textbf{62}, 16536} (2000)





\bibitem{karasiev} V.V.~Karasiev, T.~Sjostrom, J.~Dufty and S.B.~Trickey, Accurate Homogeneous Electron Gas Exchange-Correlation Free Energy for Local Spin-Density Calculations, \href{ http://link.aps.org/doi/10.1103/PhysRevLett.112.076403 }{ \textit{Phys. Rev. Lett.} \textbf{112}, 076403} (2014)






\bibitem{brown} E.W.~Brown, B.K.~Clark, J.L.~DuBois and D.M.~Ceperley, Path-Integral Monte Carlo Simulation of the Warm Dense Homogeneous Electron Gas, \href{ http://link.aps.org/doi/10.1103/PhysRevLett.110.146405 }{ \textit{Phys. Rev. Lett.} \textbf{110}, 146405} (2013)




\bibitem{node} D.M.~Ceperley, Fermion Nodes, \href{http://link.springer.com/10.1007/BF01030009}{\textit{J. Stat. Phys.} \textbf{63}, 1237-1267} (1991)





\bibitem{ebeling_richert81} W.~Ebeling, W.~Richert, and W.-D.~Kraeft,
Pad\'e Approximations for the Thermodynamic Functions of Weakly Interacting Coulombic Quantum Systems, \href{ http://onlinelibrary.wiley.com/doi/10.1002/pssb.2221040120/abstract}{\textit{phys.~stat.~sol.~(b)} {\bf 104}, 193-202} (1981)

\bibitem{ebeling_ebeling_richert82} W.~Ebeling, and W.~Richert, 
Thermodynamic Functions of Nonideal Hydrogen Plasmas, \href{ http://onlinelibrary.wiley.com/doi/10.1002/andp.19824940508/abstract }{\textit{Annalen der Physik} (Leipzig) {\bf 39}, 362-370} (1982)

\bibitem{ebeling_richert85_1} W.~Ebeling, and W.~Richert, 
Thermodynamic Properties of Liquid Hydrogen Metal, \href{ http://onlinelibrary.wiley.com/doi/10.1002/pssb.2221280211/abstract }{\textit{phys.~stat.~sol.~(b)} {\bf 128}, 467-474} (1985)

\bibitem{ebeling_richert85_2} W.~Ebeling, and W.~Richert, 
Plasma Phase-Transition in Hydrogen, \href{http://www.sciencedirect.com/science/article/pii/0375960185905213}{\textit{Phys.~Lett.~A} {\bf 108}, 80-82} (1985)





















\bibitem{ebeling2} W.~Ebeling, Free Energy and Ionization in Dense Plasmas of the Light Elements, \href{ http://onlinelibrary.wiley.com/doi/10.1002/ctpp.2150300502/abstract}{ \textit{Contrib.~Plasma Phys.}~\textbf{30}, 553 } (1990)





\bibitem{dewitt} H.E.~DeWitt, Statistical Mechanics of High‐Temperature Quantum Plasmas Beyond the Ring Approximation, \href{ http://scitation.aip.org/content/aip/journal/jmp/7/4/10.1063/1.1704974;jsessionid=8AjO6DUAODnHhlQ1CPoUsjbY.x-aip-live-03 }{ \textit{J.~Math.~Phys.}~\textbf{7}, 616} (1966)






\bibitem{stolz_ebeling} New Pad\'e approximations for the free charges in two-component strongly coupled plasmas based on the Uns\"old-Berlin-Montroll asymptotics, \href{ http://www.sciencedirect.com/science/article/pii/S0375960198006598 }{ \textit{Phys.~Lett.~A} \textbf{248}, 242} (1998)



\bibitem{stls85} S.~Tanaka, S.~Mitake, and S.~Ichimaru, Parametrized equation of state for electron liquids in the Singwi-Tosi-Land-Sj\"olander approximation, \href{ http://link.aps.org/doi/10.1103/PhysRevA.32.1896 }{ \textit{Phys.~Rev.~A} \textbf{32}, 1896} (1985)

\bibitem{stls} S.~Tanaka and S.~Ichimaru, Thermodynamics and Correlational Properties of Finite-Temperature Electron Liquids in the Singwi-Tosi-Land-Sj\"olander Approximation, \href{ http://journals.jps.jp/doi/abs/10.1143/JPSJ.55.2278 }{ \textit{J.~Phys.~Soc.~Jpn.}~\textbf{55}, 2278-2289} (1986)


\bibitem{ichimaru_rev_2} S.~Ichimaru, H.~Iyetomi, and S.~Tanaka, Statistical physics of dense plasmas: Thermodynamics, transport coefficients and dynamic correlations, \href{ http://www.sciencedirect.com/science/article/pii/0370157387901256 }{ \textit{Physics Reports}~\textbf{149}, 91} (1987)

\bibitem{ichimaru_rev} S.~Ichimaru, Nuclear fusion in dense plasmas, \href{http://link.aps.org/doi/10.1103/RevModPhys.65.255}{\textit{Rev.~Mod.~Phys.}~\textbf{65}, 255} (1993)



\bibitem{pdw84} F.~Perrot, M.W.C.~Dharma-wardana, Exchange and correlation potentials for electron-ion systems at finite temperatures, \href{ http://link.aps.org/doi/10.1103/PhysRevA.30.2619 }{ \textit{Phys.~Rev.~A}  \textbf{30}, 2619} (1984)






\bibitem{bohm} H.~Schweng and H. B\"ohm, Finite-temperature electron correlations in the framework of a dynamic local-field correction, \href{ http://link.aps.org/doi/10.1103/PhysRevB.48.2037 }{ \textit{Phys.~Rev.~B} \textbf{48}, 2037 } (1993)




\bibitem{stls2} T.~Sjostrom, and J.~Dufty, Uniform Electron Gas at Finite Temperatures, \href{ http://link.aps.org/doi/10.1103/PhysRevB.88.115123 }{ \textit{Phys.~Rev.~B} \textbf{88}, 115123} (2013)






\bibitem{stolzmann} W.~Stolzmann and M.~R\"osler, Static Local-Field Corrected Dielectric and Thermodynamic Functions, \href{ http://onlinelibrary.wiley.com/doi/10.1002/1521-3986(200103)41:2/3<203::AID-CTPP203>3.0.CO;2-S/abstract }{ \textit{Contrib.~Plasma Phys.}~\textbf{41}, 203} (2001)












\bibitem{brown3} E.W.~Brown, J.L.~DuBois, M.~Holzmann and D.M.~Ceperley, Exchange-correlation energy for the three-dimensional homogeneous electron gas at arbitrary temperature, \href{ http://link.aps.org/doi/10.1103/PhysRevB.88.081102 }{ \textit{Phys. Rev. B.} \textbf{88}, 081102(R)} (2013)



\bibitem{dufty_dutta1}
James Dufty and Sandipan Dutta, Classical representation of a quantum system at equilibrium: Theory,
\href{ http://link.aps.org/doi/10.1103/PhysRevE.87.032101 }{\textit{Phys.~Rev.~E} {\bf 87}, 032101 }(2013)

\bibitem{dufty_dutta2}
Sandipan Dutta and James Dufty, Uniform electron gas at warm, dense matter conditions,
\href{ http://iopscience.iop.org/0295-5075/102/6/67005 }{\textit{EPL} {\bf 102}, 67005} (2013)


\bibitem{spink} G.G.~Spink, R.J.~Needs, and N.D.~Drummond, Quantum Monte Carlo study of the three-dimensional spin-polarized homogeneous electron gas, \href{ http://link.aps.org/doi/10.1103/PhysRevB.88.085121 }{ \textit{Phys.~Rev.~B} \textbf{88}, 085121} (2013)









\bibitem{manfredi} N. Crouseilles, P.-A. Hervieux, G. Manfredi, Quantum hydrodynamic model for the nonlinear electron dynamics in thin metal films, \href{http://link.aps.org/doi/10.1103/PhysRevB.78.155412}{
\textit{Phys.~Rev.~B}  {\bf 78}, 155412 } (2008)

\bibitem{michta} D. Michta, F. Graziani, and M. Bonitz, 
Quantum Hydrodynamics for Plasmas--A Thomas-Fermi Theory Perspective, \href{ http://onlinelibrary.wiley.com/doi/10.1002/ctpp.201500024/abstract}{
\textit{Contrib.~Plasma Phys.}~{\bf 55}, 437} (2015)


































\end{thebibliography}
\end{document}